\documentclass[footinbib,amsfonts,dvipsnames,floatfix,twocolumn, aps]{revtex4-2}
\usepackage[T1]{fontenc} 
\usepackage{mathtools} 
\usepackage{amssymb,bm,amsthm} 
\usepackage[all]{xy} 
\usepackage{mathbbol} 
\usepackage{braket} 
\usepackage[normalem]{ulem} 
\usepackage{tikz-cd} 
\usepackage{tikz}
\usetikzlibrary{arrows.meta, positioning, shapes.geometric} 
\usepackage{todonotes} 
\usepackage{mathtools} 
\usepackage{tabularx}

\begin{document}

\title{Atomtronic routing of dipolar bosons in a four-well star potential}

\author{Karin Wittmann W.$^{1}$}
\email{karin.wittmann@ufrgs.br}

\author{Leandro H. Ymai$^{2}$}
\email{leandroymai@unipampa.edu.br}

\author{Genessi Sá Neto$^{1}$}

\author{Angela Foerster$^{1}$}

\affiliation{$^{1}$Instituto de F\'{i}sica, Universidade Federal do Rio Grande do Sul, Porto Alegre, RS, Brazil\\}
\affiliation{$^{2}$Campus Bagé, Universidade Federal do Pampa, Bagé, RS, Brazil}


\begin{abstract}
The ability to precisely control and predict the evolution of quantum states is a fundamental requirement for advancing quantum technologies. Here, we develop tunable atomic routing protocols based on an integrable model of dipolar bosons confined in a four-well potential with a star-shaped configuration. By adjusting the system parameters, we identify a harmonic dynamical regime of the atomic population that can be treated analytically,  providing a complete description of the system's behaviour for precise manipulation. We demonstrate three independent modes of control over the atomic population dynamics under the action of an external field: frequency tuning via variation in the field intensity, directional switching via spatial displacement of the field, and amplitude modulation by varying its duration. These modes operate under two distinct configurations: one source and two drains, and, in reverse order, two sources and one drain. These cases emulate an atomic 1:2 demultiplexer and 2:1 multiplexer, respectively. Our results may contribute to the development of control mechanisms in the design of quantum devices.
\end{abstract}

\maketitle


\section{Introduction}

The ability to generate and transfer quantum states between different parts of a multicomponent quantum system with high control is a key resource for advancing quantum technologies~\cite{Yuan2015,amico2021roadmap,Lee2022,luan2025}. In particular, the possibility to route and multiplex quantum modes has been explored in several research platforms, including superconducting circuits~\cite{Christensen2020, Wang2021}, cold atoms~\cite{Amico2023, Garaot2013}, and other architectures based on light-matter interactions~\cite{Parniak2017, Vernaz-Gris2018}, enabling applications in quantum information processing, atomtronics, quantum memories, and quantum networks~\cite{Cirac1997, Kimble2008, Reiserer2015, Covey2023, Li2024}.

Among these platforms, systems based on ultracold atoms stand out for their exceptional tunability and coherence properties.  When confined in an optical lattice, they offer a wide range of configurations in a periodic environment, where the coherent superposition of atoms can be controlled via laser light with high precision~\cite{Bloch2005, JAKSCH2005, Lewenstein2007, MISTAKIDIS2023}. This enables investigations ranging from few- to many-body physics, allowing the realization of useful quantum mechanical phenomena such as quantum tunneling~\cite{Albiez2005} and entanglement~\cite{Amico2008, JuliaDiaz2015}, which are essential for quantum sensing and quantum computing~\cite{Berrada2013, Evered2023}.

An additional level of control arises when considering the dipole–dipole interaction (DDI) between dipolar bosons in an optical lattice, which introduces long-range interactions in addition to the usual short-range contact interaction in ultracold gases.
This extra ingredient greatly enhances control and tunability, enabling access to a broader range of dynamical regimes and many-body quantum phase transitions~\cite{Baier2016, Goral2002, Trefzger2011, montserrat2024dipolar, MistakidisPRA2024, montserrat2025dipolar}, which are well described by the extended Bose-Hubbard model (EBHM)~\cite{Lahaye2009, Petter2019}.

So far, ultracold dipolar atoms confined in specific arrangements of a small number of wells have been observed to reach an integrable regime~\cite{Tonel2005, Wittmann2018, Grun2022} characterized by the existence of a sufficient number of conserved quantities. These conservation laws impose constraints that effectively restrict the accessible Hilbert space. The result is an orderly dynamics that can be treated analytically, providing a deeper understanding of the system, which can be highly advantageous when designing quantum devices.

In this study, we investigate the mechanisms that control information management in a system of dipolar atoms confined in a four-well potential. The system is arranged in a star configuration consisting of a central well connected to three outer wells. By adjusting the system parameters, we obtain the conditions for integrability of the system, allowing us to reach a resonant tunneling regime, characterized by harmonic atomic population dynamics in the outer wells, which can be described by a simple equation. In this scenario, we examine how the influence of an external field can be used to implement control mechanisms in different aspects - frequency, routing, and amplitude of the atomic population dynamics. Combined, these mechanisms enable the system to emulate 2:1 multiplexing and 1:2 demultiplexing operations on the atomic flow.

The article is structured as follows: In Section II, we present the theoretical framework, analyzing the conditions for integrability and the role of conserved operators in describing the system in the resonant tunneling regime. In Section III, we discuss how the frequencies of the dynamic populations can be controlled by the intensity of an additional external field. In Section IV, we explore the system’s routing capabilities under different initial conditions, introducing symmetry breaking by displacing the external field. In Section V, we present an amplitude control protocol, enabling the implementation of atomic flow (de)multiplexing. The concluding section outlines the significance of the results. More detailed technical information is provided in the appendices.


\section{System description}
\begin{figure}  
    \centering  
                \includegraphics[width=1.\linewidth]{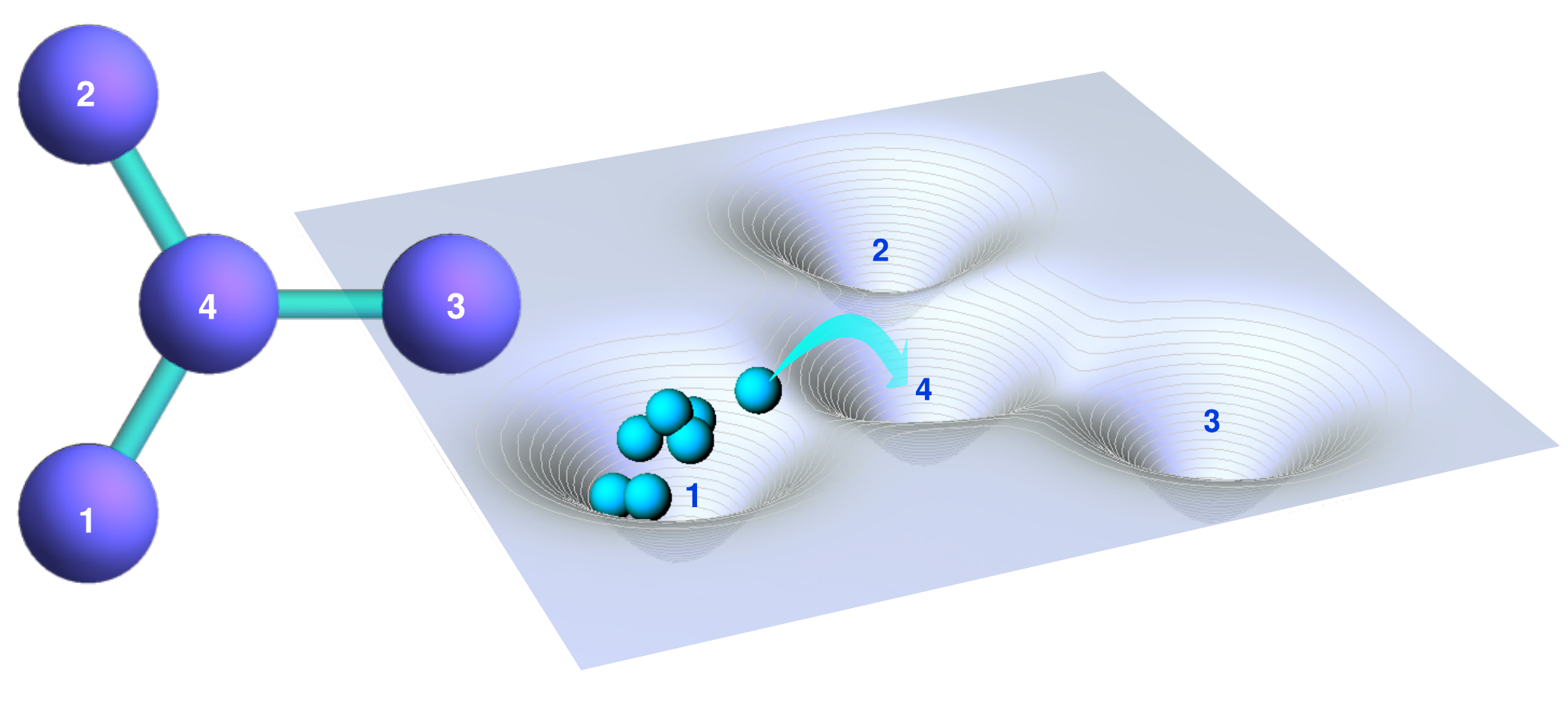} 
    \caption{Schematic representation. A central well, denoted by 4, is connected to three outer wells, denoted by 1, 2, and 3. The interactions between the atoms (blue spheres) can influence the population dynamics in the wells.}  
    \label{fig:1}  
\end{figure}

We investigate a system of ultracold dipolar bosons confined in a four-well optical trap, consisting of a central site surrounded by three outer sites, as shown in Fig. \ref{fig:1}.  

The physics of the system considered here is described by the extended Bose-Hubbard model:  
\begin{align}  
H =& \frac{U_0}{2}\sum_{i=1}^4N_i(N_i-1) +\frac{1}{2}\sum_{\substack{i,j=1\\i\neq j}}^4U_{ij}N_iN_j + \sum_{i=1}^4 \sigma_i N_i\nonumber\\  
   &-J [a_4^\dagger(a_1+a_2+a_3)+(a_1^\dagger+a_2^\dagger +a_3^\dagger)a_4],\nonumber  
\end{align}  
where \( a_i \) (\( a_i^\dagger \)) are the bosonic annihilation (creation) operators, and \( N_i = a_i^\dagger a_i \) is the number operator, which accounts for the atomic occupation at site \( i \) (\( i = 1,2,3,4 \)). The index $4$ refers to the central well. The total number of atoms in the system, \( N = N_1 + N_2 + N_3 + N_4 \), is conserved.
The parameter \( U_0 \) represents the short-range interaction energy arising from contact and dipole-dipole interactions (DDI) between atoms within the same site, while \( U_{ij} = U_{ji} \) accounts for the long-range interaction energy due to DDI between dipoles at different sites. 
The local energy gradient \( \sigma_i \) characterizes the strength of an external field at site \( i \), and \( J \) denotes the hopping rate between neighboring sites. 


{\it Integrability:} Due to the regular arrangement of the wells, the long-range interaction energies satisfy \( U_{12} = U_{23} = U_{13} \) and \( U_{14} = U_{24} = U_{34} \). We focus on the case where the long-range interaction energy between the edge sites (1, 2, and 3) is balanced with the short-range interaction energy, satisfying \( U_0 = U_{12} \).
Detailed information on explicit calculations of parameters and their tolerance to ensure integrability can be found in reference~\cite{WittmannPRA2023}. 
For the numerical simulations, we use parameters derived from dysprosium atoms ($^{164}$Dy), whose large magnetic moment allows access to the interaction regime relevant to our study.

For these parameter settings, the system enters the integrable domain~\cite{Ymai2017}.
In this case, the Hamiltonian can be written (up to a global constant) in a reduced form (see App.~\ref{app-Hamiltonian} for details):  
\begin{align}\label{h}  
H = & \, U(N_1 + N_2 + N_3 - N_4)^2 + \sigma(N_1 + N_2 + N_3 - N_4)\nonumber\\  
   & - J [a_4^\dagger (a_1 + a_2 + a_3) + (a_1^\dagger + a_2^\dagger + a_3^\dagger) a_4],  
\end{align}  
Here, \( U = (U_0 - U_{14}) / 4 \) is the effective interaction energy, and we consider the parameter \( \sigma \) representing the gradient energy between the central well and the subsystem of edge wells 1, 2, and 3 (see App.~\ref{app-Experimental} for details on the experimental implementation of this parameter).  

In the integrable regime, there are four independent conserved operators that mutually commute, equivalent to the number of degrees of freedom of the system
\footnote{In reference~\cite{Lachlan2024}, it was pointed out that the system is, in fact, superintegrable, having more conserved operators than the number of degrees of freedom of the system, although the additional conserved operators are not mutually commutative. These extra conserved operators exhibit a pseudo-spin structure, which is not relevant to our problem.}
: \( H \), \( N \), and two additional independent charges, given by  
\begin{align}  
Q = & \frac{1}{2}(N_1 + N_2-a_1^{\dagger} a_2 - a_2^{\dagger} a_1) \\  
\widetilde{Q} = & \frac{1}{6}(N_1 + N_2 + 4 N_3+a_1^{\dagger} a_2 + a_2^{\dagger} a_1) \nonumber\\
&\hspace{0.8cm} - \frac{1}{3} [ a_3^{\dagger}(a_1 + a_2)  + (a_1^{\dagger}+a_2^{\dagger})a_3].\nonumber  
\end{align}  

In general, systems with fewer conserved quantities than degrees of freedom tend to exhibit chaotic dynamics and thermalize~\cite{wittmannChaos2022,wittmannChaosPRA2024,wittmann2024}. In contrast, the existence of sufficient conserved quantities imposes several restrictions on the system, favouring more regular dynamics and preventing thermalization. One of the key advantages of integrability in this system is that it allows the atomic population dynamics to exhibit resonant behavior, governed by an effective Hamiltonian that depends linearly on the conserved charges. 
As we will discuss, this simplifies the mathematical treatment of the system, even in the presence of symmetry breaking.


\vspace{0.3cm}
{\it Quantum dynamics:}
The Hamiltonian~\eqref{h} governs the time evolution of an initial state \( |\Psi_{\rm ini}\rangle \), according to the equation  
\[
|\Psi(t)\rangle = e^{-iHt} |\Psi_{\rm ini}\rangle
\]
where we set \(\hbar=1\). The above state is represented in the Hilbert space using the Fock basis:  
\begin{eqnarray}
|n_1,\,n_2,\,n_3,\,n_4\rangle = \frac{(a_1^\dagger)^{n_1}}{\sqrt{n_1!}} \frac{(a_2^\dagger)^{n_2}}{\sqrt{n_2!}} \frac{(a_3^\dagger)^{n_3}}{\sqrt{n_3!}} \frac{(a_4^\dagger)^{n_4}}{\sqrt{n_4!}} |0,0,0,0\rangle,\nonumber
\end{eqnarray}
with a fixed total number of atoms \( N=n_1+n_2+n_3+n_4 \). Then, the dynamics of the atomic population in the wells are described by the expectation values  
\[
\langle N_j\rangle = \langle \Psi(t)|N_j|\Psi(t)\rangle,
\]
for \( j=1,2,3,4 \).

\vspace{0.3cm}

{\it Resonant tunnelling regime:}
We focus on the resonant tunneling regime, characterized by the condition \( |U(N-2n_4)+\sigma/2|/J \gg 1 \). In this regime, the populations of the edge sites exhibit coherent harmonic oscillations induced by a second-order process~\cite{Lahaye2010}, while the number of atoms in the central well remains conserved. 
For the case $\sigma=0$, this behaviour results from two combined effects.
First, for sufficiently large values of $U$, the first term of Hamiltonian introduces a `quadratic tilt' between the central well and the subsystem of outer wells. This tilt tends to confine the atoms in these subsystems, preserving the number of atoms in each of them over time. On the other hand, the cancellation of the long-range interaction term $(U_{12}-U_0)(N_1N_2+N_2N_3 + N_1N_3)$ (see App.~\ref{app-Hamiltonian}), due to the integrability condition, transforms the outer wells into an effectively noninteracting subsystem.  As a result, atoms in the outer wells can tunnel freely among them.

\begin{figure}
    \centering
          \includegraphics[width=1.0\linewidth]{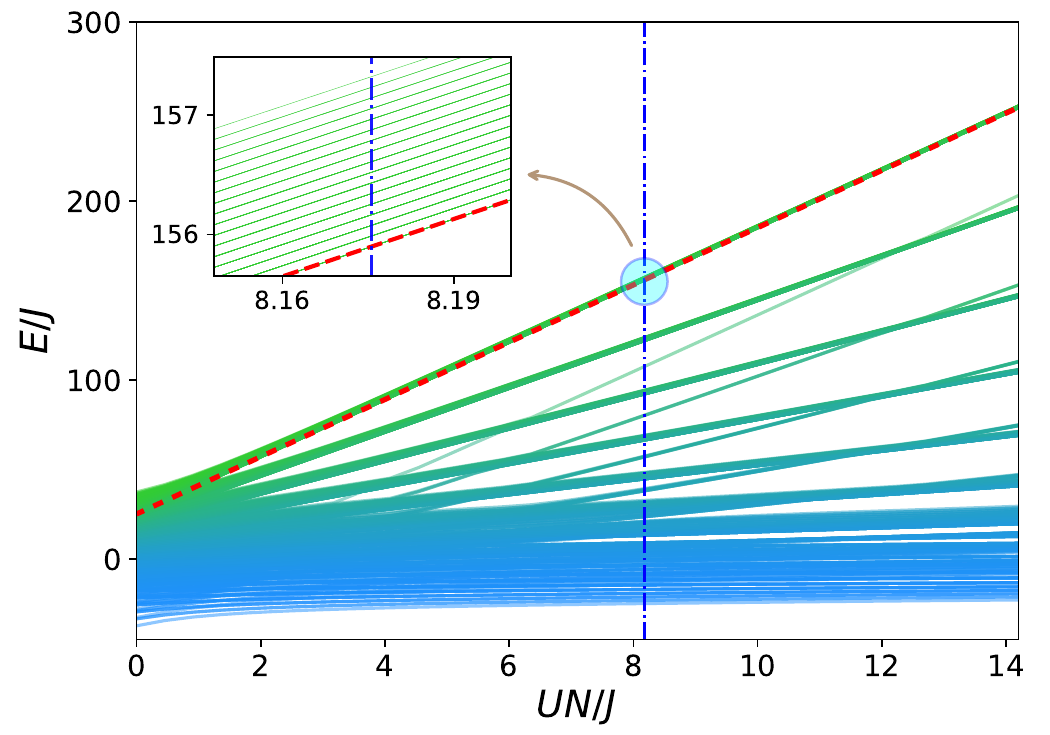}
   \caption{Energy levels as a function of \(UN/J\). 
   Plot of \(E/J\) versus \(UN/J\) for the Hamiltonian \eqref{h}, varying \(U\) for $N=16$. The vertical dashed line indicates the value of \(UN/J\), corresponding to \(U/J = 0.51\) and
    \(\sigma/J = 1.56\). 
   The red dashed line indicates the expectation value $\langle \Psi_{\rm ini}|H|\Psi_{\rm ini}\rangle$ for the initial state $|\Psi_{\rm ini}\rangle = |14,2,0,0\rangle$. The circle marks a region within this band, which is magnified in the inset plot showing the uniform distribution of energy levels.}
    \label{fig:Energias}
\end{figure}

Another key feature of the resonant regime is the formation of bands in the energy spectrum, each with uniformly spaced energy levels, as shown in Fig.~\ref{fig:Energias}. 
Only the energy levels in the band close to the expectation value of the initial state's energy effectively contribute to the dynamics. These, in turn, exhibit harmonic behavior due to the regular energy level structure, analogous to that of a harmonic oscillator. This behavior can be described by an effective Hamiltonian (see App.~\ref{app-Heff}):  
\begin{eqnarray}
H^{\rm eff} &=& \mathcal{J}_{\rm eff}[2(N_1+N_2+N_3)-3(Q+\widetilde{Q})],\label{heff}
\end{eqnarray}  
where $ \mathcal{J}_{\rm eff}\equiv \mathcal{J}(0)$ is the effective hopping rate, 
and we define the function
\begin{eqnarray}\label{Jx}
\mathcal{J}(x) = \frac{J^2[4U(N+1)+2\sigma +x]}{[4U(N-2n_4)+2\sigma+x]^2-(4U)^2}, \quad (x\in \mathbb{R})\nonumber\\
\end{eqnarray}
which depends on the initial population in the central well.

Using the effective Hamiltonian~\eqref{heff} in the time-evolution operator, we derive the following analytical expression for the expectation values of the populations at the edge sites for the initial state $|\Psi_{\rm ini}\rangle = |n_1,n_2,n_3,n_4\rangle$ (see App.~\ref{app-ExpValue} for details):  
\begin{eqnarray}
\langle N_{j=1,2,3}\rangle &=& n_j+\frac{4}{9}(N-3n_j-n_4)\,\sin^2\left(\frac{3}{2}\mathcal{J}_{\rm eff}\,t\right), \quad\,
\label{n_k}
\end{eqnarray}
while the population of the central well remains constant, $\langle N_4\rangle = n_4$. 
From now on, we will consider $n_4=0$ in numerical simulations for simplicity. 

The harmonic behavior, described analytically by Eq.~\eqref{n_k}, highlights the potential for controlled atomic transport in the resonant regime. A key advantage of the proposed setup is the ability to implement multiple levels of control over the atomic dynamics: the frequency of oscillations, the direction of atomic flow, and the amplitude of population transfer can all be independently tuned through external fields, as will be detailed in the following sections.


\section{Frequency control}
In this section, we focus on frequency control, which is achieved by varying the energy gradient $\sigma$. This gradient directly affects the analytical expression for the populations (Eq.~\eqref{n_k}), enabling precise and predictable modulation of the oscillation period. Such control is essential for timing and synchronization in atomtronic applications. We now analyze how the oscillation frequency depends on $\sigma$ and evaluate the fidelity of the dynamics to identify the parameter regimes where resonant behavior persists.

\begin{figure}[!ht] 
\center
\includegraphics[width=1\linewidth]{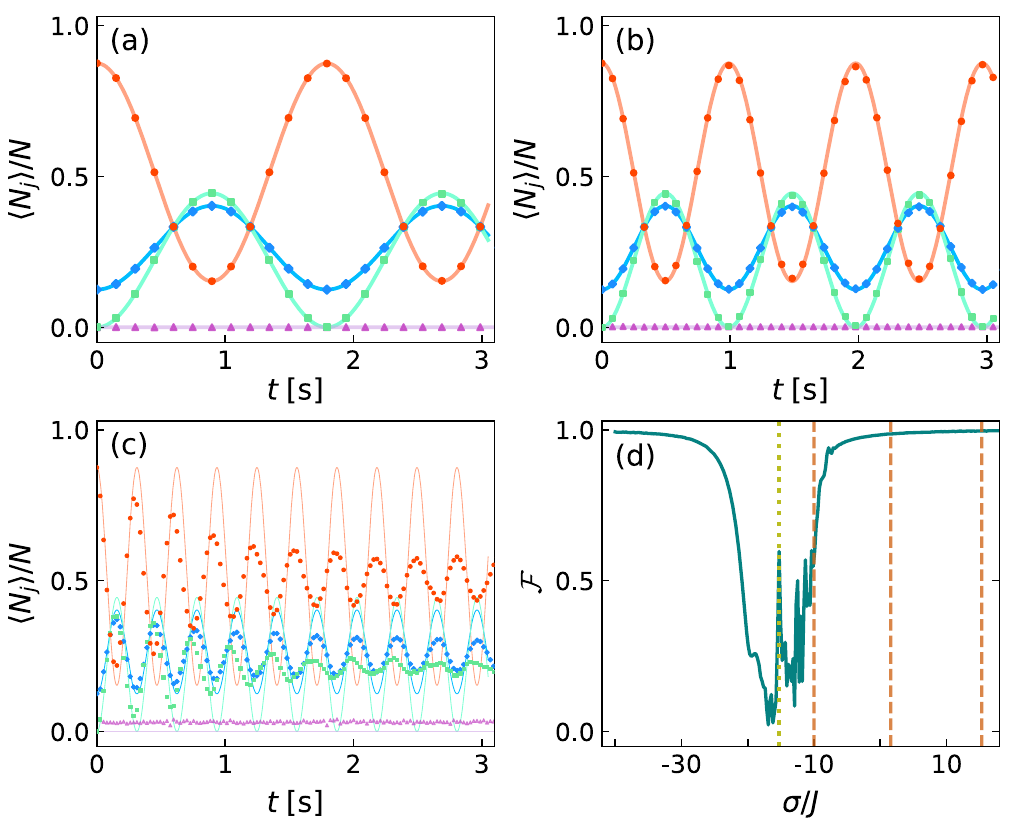}
\caption{Quantum dynamics. Expectation values of the fractional populations $\langle N_1\rangle/N$ (red), $\langle N_2\rangle/N$ (blue), $\langle N_3\rangle/N$ (green), and $\langle N_4\rangle/N$ (purple). Numerical results obtained using the Hamiltonian~\eqref{h} (markers) are compared with the analytical expression in Eq.~\eqref{n_k} (solid lines), for parameter values (a) $\sigma/J = 15.3$, (b) $\sigma/J = 1.57$, and (c) $\sigma/J = -10$. (d) Fidelity $\mathcal{F}$. The vertical dotted line represents the critical value $\sigma/J = \sigma_{\rm crit}/J = -15.3$, and the vertical dashed lines represent the values $\sigma/J = -10,\, 1.57$, and $15.3$. In all cases we consider $U/J = 0.51$ and the initial state $|\Psi_{\rm ini}\rangle = |14,2,0,0\rangle$.}
\label{fig:frequency}
\end{figure}

For sufficiently large values of $|\sigma /J|$, the external field induces an energy gradient between the central well and the others, favoring the confinement of atoms within these two subsystems and enhancing the resonant behavior of the atomic population even in the weak-interaction regime, $|U/J| < 1$. This implies that the oscillation period of the populations can be reduced and, more importantly, that the oscillation frequency can be controlled by the intensity of the external field. An example of such control is shown in Fig.~\ref{fig:frequency}(a)-(b), where we compare the dynamics of the fractional populations for different values of $\sigma$, demonstrating good agreement between the analytical expression~\eqref{n_k} and the numerical simulations based on the Hamiltonian~\eqref{h}. We observe that the oscillation frequency increases as $\sigma$ approaches the critical value $\sigma_{\rm crit}=2U(1-N)$, where the parameter $\mathcal{J}_{\rm eff}$ diverges (see Eq.~\eqref{Jx}). However, frequencies cannot be increased arbitrarily: as $\sigma$ approaches $\sigma_{\rm crit}$, the system is gradually enters an off-resonant regime, as illustrated in Fig.~\ref{fig:frequency}(c).

To identify the resonant regime as the parameter $\sigma$ changes, we investigate how the system returns to its initial configuration after a period of $2\pi/3\mathcal{J}_{\rm eff}$, by evaluating the fidelity, defined as $\mathcal{F}=|\langle \Psi_{\rm in}|\Psi(2\pi/3\mathcal{J}_{\rm eff})\rangle|$. Fig.~\ref{fig:frequency}(d) shows how the fidelity varies with $\sigma/J$, indicating the resonant regime when $\mathcal{F} \approx 1$.


\section{Routing control}\label{routing}
{\it Symmetry breaking:} 
We now investigate the mechanism that controls the tunneling processes within the subsystem of outer sites $i \in \{1,2,3\}$. 
To this end, an asymmetry is introduced among the outer wells via an additional harmonic potential 
generated by an external field slightly displaced toward one of the three outer wells. This asymmetry enables control over the direction of atomic flow within the subsystem.

\begin{figure}[h!]
\center
\includegraphics[scale=1]{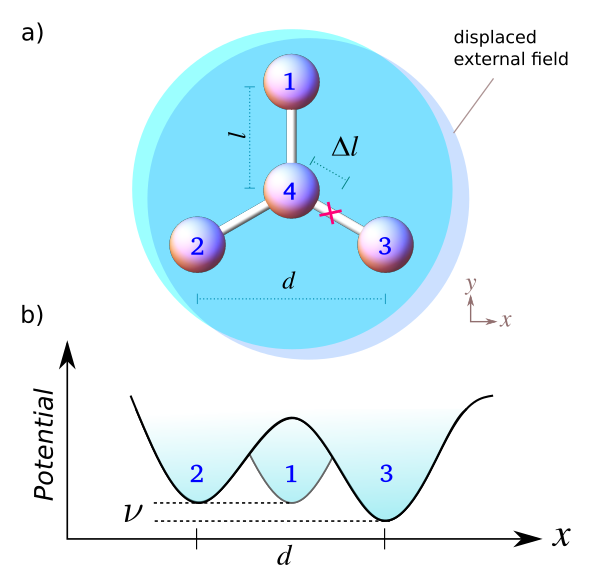}
\caption{
(a) Schematic representation of the external field displacement. An additional harmonic potential (gray disk) is displaced from well 4 toward well 3 by a distance $\Delta l$ and centered at the red cross. (b) Symmetry breaking. The displaced external field induces an energy offset $\nu$ between well 3 and the other two, which are kept at the same level. As a result, wells 1 and 2 form a resonant tunnel channel, denoted by $C_{1,2}$.}
\label{fig:break1}
\end{figure}

Mathematically, the external field introduces an additional term that breaks the symmetry of the Hamiltonian \eqref{h} by permuting the indices of wells 1, 2, and 3 modifying it as follows: 
\begin{eqnarray}
H_{k} = H + \nu (N_i + N_j - 2N_k), \quad (i\neq j\neq k), \label{hi}
\end{eqnarray}  
where \(H\) is the Hamiltonian given in Eq.~\eqref{h}, $k \in\{ 1, 2, 3\}$, denotes the well toward which the external field is displaced, and $i$ and $j$ are the other two wells different from $k$. The parameter \(\nu \propto \Delta l\) characterizes the strength of the external field displaced from the center toward well $k$ by $\Delta l$. For brevity, we refer to this field as \(F_k\).
The remaining two wells $i$ and $j$ are kept at the same potential depth, forming a tunneling channel denoted by $C_{i,j}$. For instance, when $k=3$, the Hamiltonian becomes
\begin{eqnarray}
H_3 = H+\nu(N_1+N_2-2N_3),\nonumber
\end{eqnarray}
due to the applied field $F_3$, so that wells 1 and 2 form a tunneling channel $C_{1,2}$. A schematic illustration of the external field displacement is shown in Fig.~\ref{fig:break1}, and further experimental details are provided in App.~\ref{app-Experimental}.

\noindent 
Under the action of the field $F_k$, 
the population of the central site remains constant over time, $\langle N_4\rangle_k = \langle \Psi_k(t)|N_4|\Psi_k(t)\rangle = n_4$ for all $k = 1, 2, 3$, and the time evolution of the system is described by the state $|\Psi_k(t)\rangle = e^{-i H_k t}|\Psi_{\rm ini}\rangle$. By applying the corresponding effective Hamiltonian \eqref{hi} in the time-evolution operator, we obtain an expression for the populations of the outer wells (see Apps.~\ref{app-Heff} and \ref{app-ExpValue} for details):
\begin{eqnarray}
\langle N_j\rangle_{k} = n_j + (n_{6-j-k} - n_j)(1-\delta_{j,k})\sin^2(\zeta  t),
\label{njk2}
\end{eqnarray}
where $\delta_{j,k}$ is the Kronecker delta and $\zeta$ is the effective frequency, given by
\begin{eqnarray}
\zeta &=&\mathcal{J}(\nu)\left[1+\frac{\mathcal{J}(\nu)}{3\nu}\right],\label{zeta}
\end{eqnarray}
with $\mathcal{J}(\nu)$ defined in Eq.~\eqref{Jx}. 
In what follows, we explore the dynamics governed by \( H_k \) and show how the field \( F_k \) enables control over the tunneling process and the transport of quantum states in the subsystem of wells 1, 2, and 3.

{\it Case A: 1 source and 2 drains -} Under the influence of the external field \( F_k \) described above, the system can act as a directional switching device, with one source (well 1) and two drains (wells 2 and 3). The field \( F_k \),  for $k=$ 2 or 3, determines which tunneling channel ($C_{1,3}$ or $C_{1,2}$, respectively) is activated, enabling atoms to flow periodically from the source to one of the drains.

\begin{figure}[t]
\center
\includegraphics[width=1\linewidth]{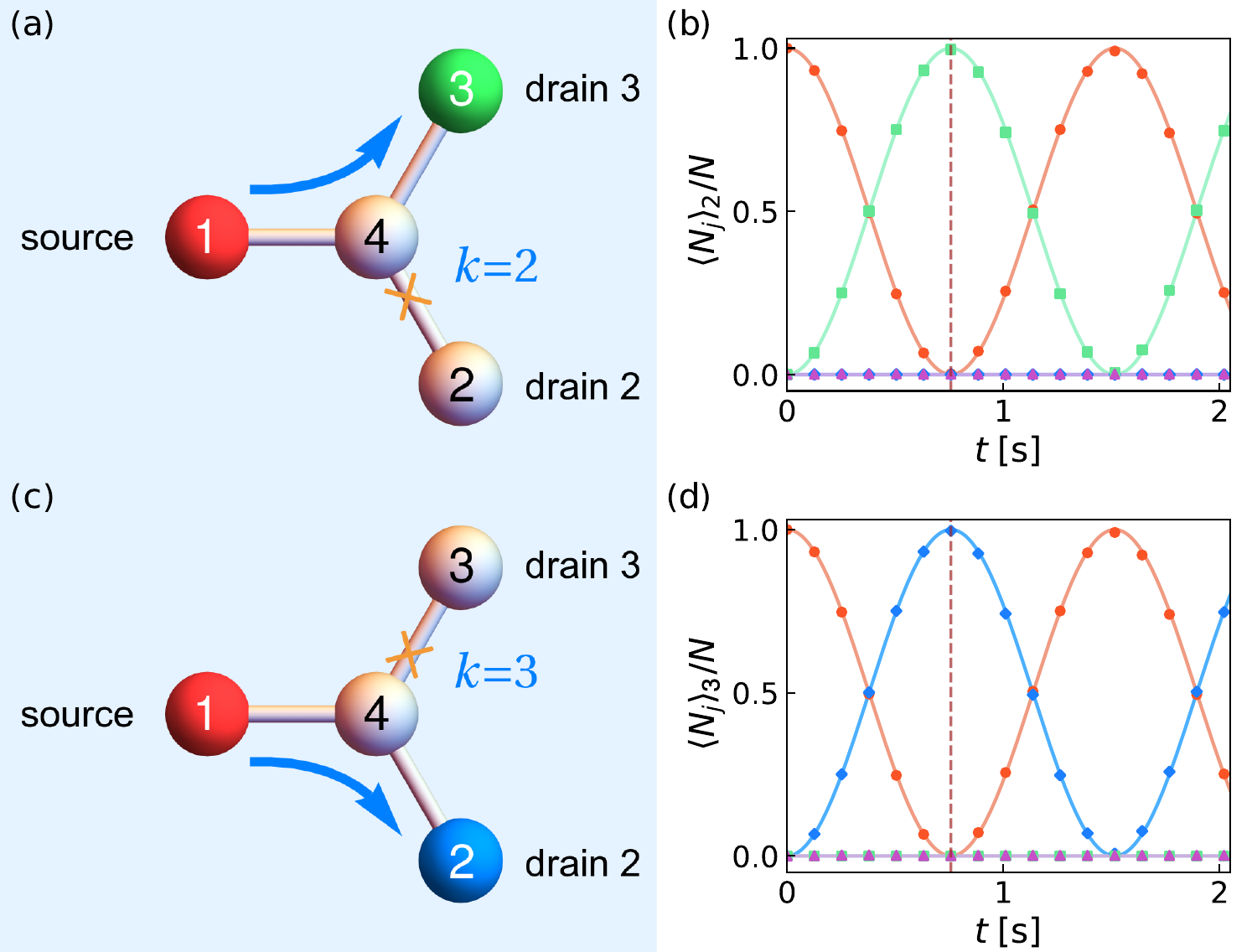}
\caption{
Schematic representations of the directional control mechanism, (a) and (c). The symbol `$\times$' indicates the center of the external field $F_2 (F_3)$, which selects the tunneling channel through which the atomic flow (blue arrow) from the source - well 1 - toward the drain - well 3 (2) - occurs.  Corresponding dynamics are shown in (b) and (d). Fractional populations: $\langle N_1\rangle_k/N$ (red), $\langle N_2\rangle_k/N$ (blue), $\langle N_3\rangle_k/N$ (green), and $\langle N_4\rangle_k/N$ (purple). Comparison between analytical expression~\eqref{njk2} (solid lines) and the numerical results using  Hamiltonian~\eqref{hi} (markers), for $k=2$ (b) and $k=3$ (d). In both cases, the parameters are: $U/J = 0.51$, $\sigma/J = 1.56$, $\nu/J = 1.05$, and initial state $|\Psi_{\rm ini}\rangle = |16,0,0,0\rangle$. The transfer time $t=\tau$ of the atoms from the source to the drain is indicated by the dashed vertical line.}
\label{fig:directional}
\end{figure}

Figure~\ref{fig:directional} shows a schematic of the directional switching protocol together with the corresponding quantum dynamics, comparing numerical simulations based on the Hamiltonian~\eqref{hi} with the analytical expression~\eqref{njk2}, for the initial state \( |\Psi_{\rm ini}\rangle = |16,0,0,0\rangle \). 
The applied field \(F_2\) (Figs.~\ref{fig:directional}(a--b)) or \(F_3\) (Figs.~\ref{fig:directional}(c--d)) selects the tunneling channel \(C_{1,3}\) or \(C_{1,2}\), respectively, along which the atoms oscillate with Rabi frequency \(2\zeta\). For \(k=1\), Eq.~\eqref{njk2} gives \(\langle N_1\rangle_1 = n_1\), so the atoms remain confined in the source without tunneling to the drains.

By operating the system over a time interval
\begin{eqnarray}
\tau = \frac{\pi}{2 |\zeta |},\nonumber
\end{eqnarray}
one can implement a simple atom routing protocol that fully transfers the population from the source to a selected drain. This is demonstrated by setting $n_2 = n_3 = 0$ in Eq.~\eqref{njk2} and evaluating it at $t = \tau$, which gives:
\begin{eqnarray}\label{eq:analitic}
\langle N_2\rangle_{k} = n_1\delta_{k,3}, \qquad \langle N_3\rangle_{k} = n_1\delta_{k,2}.
\end{eqnarray}

These expressions show that all $n_1$ atoms initially in site 1 are completely transferred to site $3$ $(2)$ under the action of the field $F_2$ $(F_3)$, as illustrated in Fig. 5(b) and (d).
This routing mechanism is schematically represented as:

\[
\centering
\begin{tabularx}{0.8\linewidth}{l c}
\hline
\qquad\qquad $t = 0$ & \hspace{2.cm} $t = \tau$ \\
\hline
\end{tabularx}
\]

\vspace{-0.8cm}
\[
\xymatrix  @R=1.5mm @C=0.5cm @M=2pt {
                    &                             & |0,0,n_1,n_4\rangle \\
|n_1,0,0,n_4\rangle \ar[drr]_{k=3} \ar[urr]^{k=2} & &\\
                    &                             & |0,n_1,0,n_4\rangle }
\]


{\it Case B: 2 sources and 1 drain –} We now analyze the case where atoms are initially distributed across two sources (wells 1 and 2), and the applied field selects one of them to route the atoms to a common drain (well 3). At time $t = \tau$, for initial Fock state $|\Psi_{\rm ini}\rangle = |n_1, n_2, 0, n_4\rangle$, the expectation value from equation~\eqref{njk2} yields:
\begin{eqnarray}
\langle N_3\rangle_{k} &=& n_{3-k}. \nonumber
\end{eqnarray}

This result shows that the field $F_1(F_2)$, selects the $n_2(n_1)$ atoms initially located in well 2(1) to be transferred to well 3 after a time interval $\tau$, while the unselected set of atoms remains trapped. The routing process is schematically represented as:

\[
\centering
\begin{tabularx}{0.8\linewidth}{l c}
\hline
\qquad\quad $t = 0$ & \hspace{2.5cm} $t = \tau$ \\
\hline
\end{tabularx}
\]

\begin{figure}[!ht]
\center
  \includegraphics[width=1\linewidth]{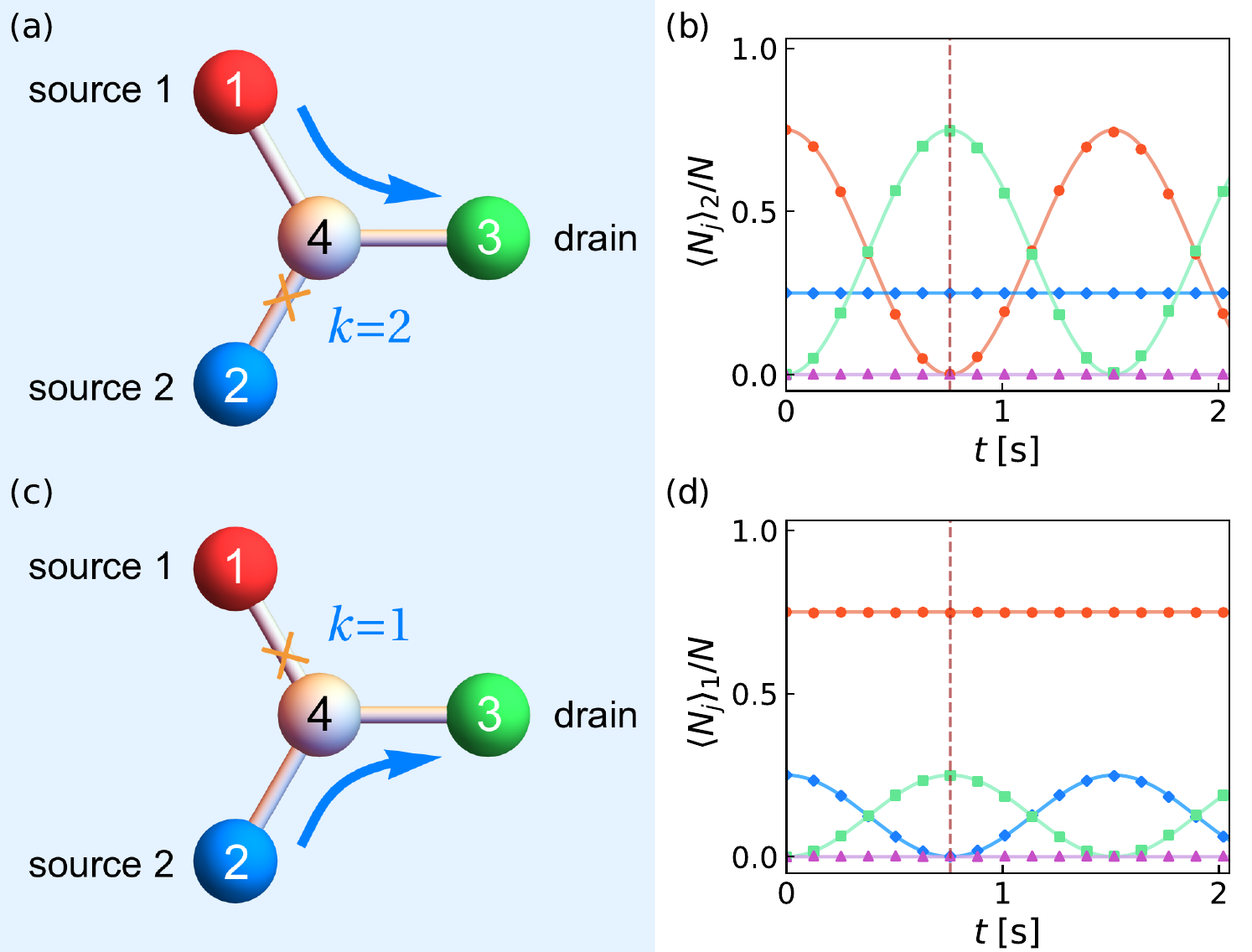}
\caption{  
Schematic of the atomic selector mechanism, in (a) and (c). The symbol `$\times$' marks the center of the external field $F_{k=1,2}$, which acts as a selector for the atomic flux (blue arrow). Panels (b) and (d) display the corresponding time evolution of the fractional populations $\langle N_1 \rangle_k / N$ (red), $\langle N_2 \rangle_k / N$ (blue), $\langle N_3 \rangle_k / N$ (green) and $\langle N_4\rangle_k/N$ (purple) for $k = 2$ and $k = 1$, respectively, comparing analytical results from Eq.~\eqref{njk2} (solid lines) with numerical simulations using Hamiltonian~\eqref{hi} (markers). Vertical dashed lines indicate $t = \tau$. In both cases, the parameters are:  $U/J = 0.51$, $\sigma/J = 1.56$, $\nu/J = 1.05$, and $|\Psi_{\rm ini}\rangle = |12,4,0,0\rangle$.}

\label{fig:diplexer}
\end{figure}

\vspace{-0.7cm}
\[\xymatrix @R=0.5mm @C=1.5cm @M=2pt { 
 & |0,n_2,n_1,n_4\rangle \\
|n_1,n_2,0,n_4\rangle  \ar[dr]_{k=1} \ar[ur]^{k=2} &  \\
& |n_1,0,n_2,n_4\rangle }
\]
Figure ~\ref{fig:diplexer} presents a schematic of this protocol and the corresponding quantum dynamics of the fractional populations for the initial state $|\Psi_{\rm ini}\rangle = |12,4,0,0\rangle$. We find good agreement between the results of the numerical simulation and the theoretical prediction from Eq.~\eqref{njk2}, 
which verifies the functionality of the system as an atomic flow selector, mediating between two sources and a single drain. 

Here, the amplitude of oscillation of the atomic population in the tunneling channel is strictly determined by the initial number of atoms in the sources. A protocol for the continuously adjusting of this amplitude will be discussed next.

\section{Amplitude control}\label{amplitude}
In this section, we build on the features
discussed previously by introducing a second stage in the quantum evolution, which incorporates a continuous tuning parameter. This addition enables precise control of the atomic oscillation amplitude through the tunneling channels and allows for the preparation and guided transfer of coherent states.

{\it Case A: 1 source and 2 drains -} As in previous cases, the protocol begins with the initial state \( |\Psi_{\rm ini}\rangle = |n_1,0,0,n_4\rangle \), which evolves during a time interval \( \tau \) under the action of the field \( F_{k=2,3} \), resulting in the states  
\[
|\Psi^{\rm I}_k\rangle = e^{-i\tau H_k}|\Psi_{\rm ini}\rangle = 
\begin{cases}
|0,0,n_1,n_4\rangle, & k=2 \\
|0,n_1,0,n_4\rangle, & k=3\,.
\end{cases}
\]
To control the amplitude of the populations at sites 2 and 3, we then consider an instantaneous switch of the field from \( F_{k=2,3} \) to \( F_1 \). The system subsequently evolves for a time interval \( q\tau \), with \( q \in [0,1] \) denoting the evolution time in units of $\tau$. This process is described by the following state:
\begin{eqnarray}
|\Psi_k^{\rm II}(q)\rangle &=& \exp(-iq\tau H_1)|\Psi^{\rm I}_k\rangle \nonumber\\
 &=&\begin{cases}
 |\alpha,\beta\rangle\rangle_{3,2}|0,n_4\rangle_{14},& k=2\\
 |\alpha,\beta\rangle\rangle_{2,3}|0,n_4\rangle_{14},& k=3
 \end{cases}\nonumber
 \end{eqnarray}
where we define the coherent state~\cite{Byrnes2015} of sites \( i \) and \( j \) as  
\[
|\alpha,\beta\rangle\rangle_{i,j} = \frac{1}{\sqrt{n_1!}}\left(\alpha a_i^\dagger + \beta a_j^\dagger\right)^{n_1}|0,0\rangle,
\]
with the coefficients  
\[
\alpha = \cos\left(\frac{q\pi}{2}\right), \qquad \beta = e^{-i\pi/2} \sin\left(\frac{q\pi}{2}\right),
\]  
controlled by the parameter \( q \). The sequence of the states generated in the protocol steps is schematically represented below:

\[
\renewcommand{\arraystretch}{1.1} 
\begin{tabularx}{\linewidth}{l l r}
\hline
\qquad\,\,\, $t = 0$ & \qquad\qquad $t = \tau$ & \qquad\qquad $t = \tau(1+q)$ \\
\hline
\end{tabularx}
\renewcommand{\arraystretch}{1.0} 
\]
\vspace{-0.7cm}
\[\xymatrix  @R=1.5mm @C=0.5cm @M=2pt {
                    &                             |0,0,n_1,n_4\rangle  \ar[r]^-{k=1} & |\alpha,\beta\rangle\rangle_{3,2}|0,n_4\rangle_{14}\\
|n_1,0,0,n_4\rangle \ar[dr]_-{k=3} \ar[ur]^-{k=2}  & & \\
                    &                            |0,n_1,0,n_4\rangle  \ar[r]_-{k=1}& |\alpha,\beta\rangle\rangle_{2,3}|0,n_4\rangle_{14} }
\]

These final states allow for the analytical evaluation of the population imbalance between wells 2 and 3 as a function of \( q \), given by
\begin{eqnarray}
 \langle \Psi_k^{\rm II}(q)|N_2-N_3|\Psi_k^{\rm II}(q)\rangle/n_1 &=& (-1)^{k+1}\cos(q\pi).\quad
\label{imb}
\end{eqnarray}

\begin{figure}[!ht]
\center
\includegraphics[scale = 0.52]{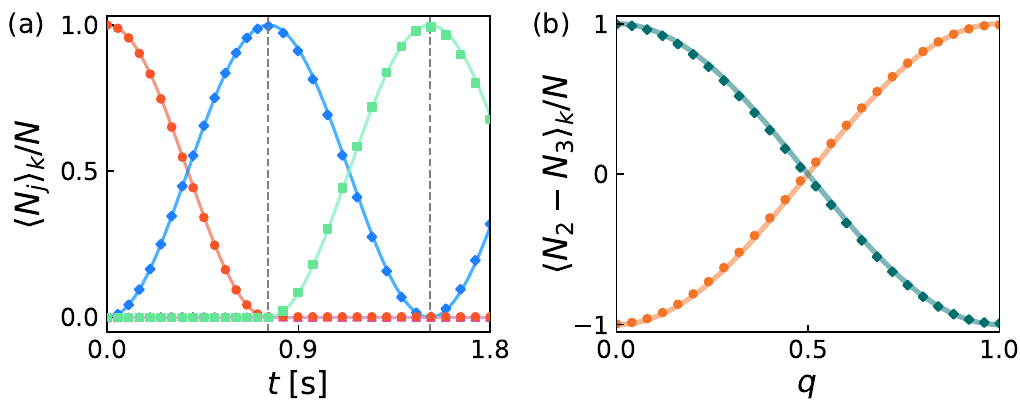}
\caption{(a) Fractional expectation values $\langle N_1\rangle_k/N$ (red), $\langle N_2\rangle_k/N$ (blue), $\langle N_3\rangle_k/N$ (green) and $\langle N_4\rangle_k/N$ (purple). The vertical dashed lines indicate $t = \tau$ and $t = 2\tau$. (b) Population imbalance of wells 2 and 3 as a function of $q$ at $t=\tau(1+q)$: numerical simulation (markers) and the analytical expression~\eqref{imb} (solid lines), for $k = 2$ (teal) and $k = 3$ (orange). We consider $U/J = 0.51$, $\sigma/J = 1.56$, $\nu/J = 1.05$, and initial state $|\Psi_{\rm ini}\rangle = |16,0,0,0\rangle$.}
\label{fig:imb}
\end{figure}

Fig.~\ref{fig:imb}(a) shows that the atomic population initially at well 1 is fully transferred to well 2 during the first step of the protocol. Then, the instantaneous change of external field at $t=\tau$ changes the direction of atomic flux toward well 3, in which the fractions of the atomic populations distributed in wells 2 and 3 can be determined by the duration of the second step of the protocol characterized by the parameter $q$. Figure~\ref{fig:imb}(b) shows the comparison between the numerical simulation of the population imbalance of wells 2 and 3 and the analytical expression~\eqref{imb} as a function of the parameter $q$, indicating a good agreement between the results.      

{\it Case B: 2 sources and 1 drain -} Now, we consider the steps of the previous case in reverse order, with two sources (wells 1 and 2) and one drain (well 3). The first stage of the protocol prepares a state with a controlled distribution of atoms in wells 1 and 2. To this end, we start from the initial Fock state \( |\Psi_{\rm ini}\rangle = |n_1,0,0,n_4\rangle \) and apply the field \( F_3 \). The system evolves over a time interval \( q\tau \), with \( q \in [0,1] \), leading to the formation of a coherent state:
\begin{eqnarray}
|\Psi^{\rm I}_3(q) \rangle &=& \exp(-iq\tau H_3)|\Psi_{\rm ini}\rangle \nonumber \\
&=& |\alpha,\beta\rangle\rangle_{1,2}|0,n_4\rangle_{3,4}, \label{cs}
\end{eqnarray}
where, for convenience, the initial time of this step is taken to be at \( t = -q\tau \). Note that for the extreme values of the parameter \( q \), (\( q = 0 \) and \( q = 1 \)), the resulting states (up to a global phase) are \( |\Psi^{\rm I}_3(0)\rangle = |n_1,0,0,n_4\rangle \) and \( |\Psi^{\rm I}_3(1)\rangle = |0,n_1,0,n_4\rangle \), respectively, indicating that all atoms are localized in one of the sources.  For intermediate values \( 0 < q < 1 \), the state in the sources is a coherent state, in which the probability distribution of atoms is determined by the coefficients $\alpha$ and $\beta$.

Once the state~\eqref{cs} has been prepared, the second step of the protocol begins after the field \( F_3 \) is instantly switched to \( F_1 \) or \( F_2 \). This stage is subsequently described by the state:
\begin{eqnarray}
|\Psi_k^{\rm II}(t)\rangle = \exp(-it H_k)|\Psi^{\rm I}_3(q)\rangle,\nonumber 
\end{eqnarray}
where $k=1$ or $k=2$. From this state, the expectation values can be calculated, yielding:
\begin{equation}
\langle\Psi_k^{\rm II}(t)|N_j|\Psi_k^{\rm II}(t)\rangle = 
\begin{cases}
A_j(q)\cos^2[(1\!-\!\delta_{j,k})\zeta  t], & j\neq3 \\[2pt]
A_{3-k}(q)\sin^2(\zeta  t), & j=3
\end{cases}
\label{ncdipl}
\end{equation}
where the amplitudes of the oscillations through the tunneling channel \( C_{3-k,3} \) are determined by the state \eqref{cs}:
\begin{eqnarray}
A_1(q) &=& \langle \Psi^{\rm I}_3(q)|N_1|\Psi^{\rm I}_3(q)\rangle =
n_1\cos^2\left(\frac{q\pi}{2}\right),\nonumber\\
A_2(q) &=& \langle \Psi^{\rm I}_3(q)|N_2|\Psi^{\rm I}_3(q)\rangle =
n_1\sin^2\left(\frac{q\pi}{2}\right).\nonumber
\end{eqnarray}
From these expressions, it is clear that
\begin{eqnarray}
A_1(q) + A_2(q) = n_1,\nonumber
\end{eqnarray}
demonstrating that the parameter \( q \) effectively tunes the oscillation amplitude in the tunneling channels. 

\begin{figure}[!ht]
\center
\includegraphics[width=1\linewidth]{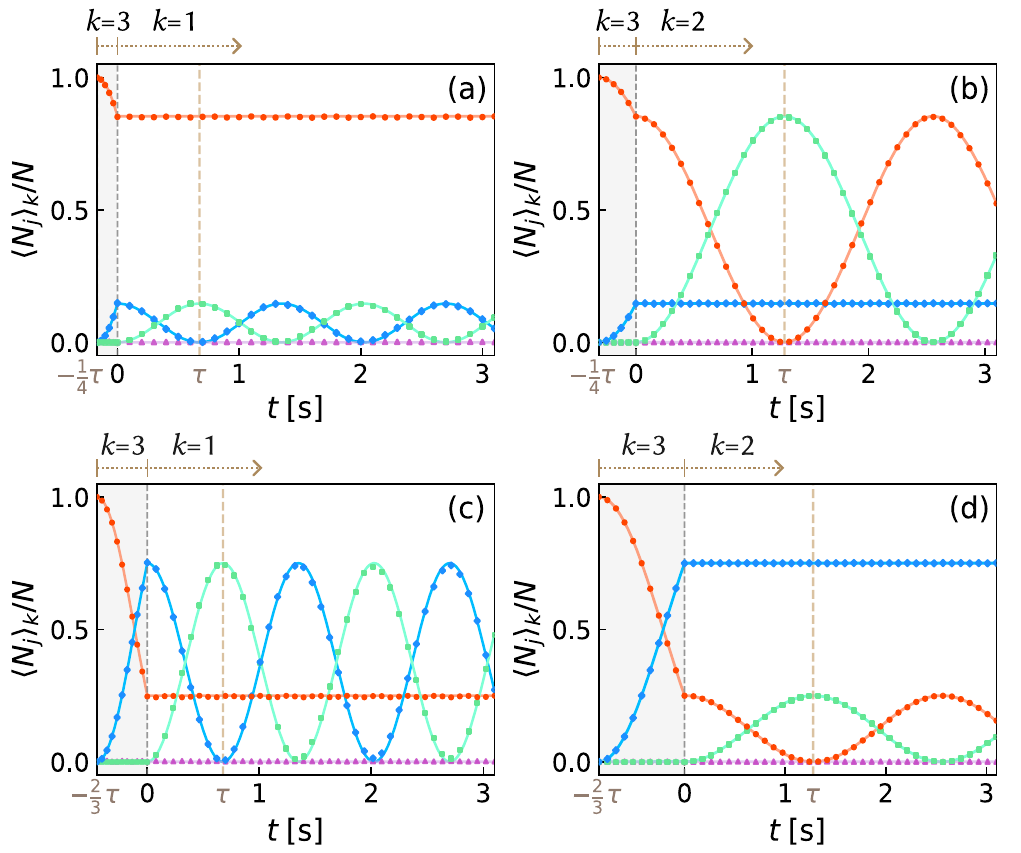}
\caption{
Time evolution of fractional populations $\langle N_1\rangle_{k}/N$ (red), $\langle N_2\rangle_{k}/N$ (blue), $\langle N_3\rangle_{k}/N$ (green), and $\langle N_4\rangle_{k}/N$ (purple), obtained from numerical simulations (markers) and from the analytical expressions \eqref{njk2} and \eqref{ncdipl} (solid lines). Parameters: $U/J = 0.51$, $\nu/J = -3.14$, and initial state $|\Psi_{\rm ini}\rangle = |16,0,0,0\rangle$. Two values of the  parameter $\sigma$ are considered:  $\sigma/J = 1.56$ (a,c) and $\sigma/J = 15.3$ (b,d). 
The shaded regions indicate the preparation of the coherent state before the second stage, which begins at $t = 0$. The first stage is driven by field $F_3$ during the time interval $q\tau$, with $q = 1/4$ in (a,b) and $q = 2/3$ in (c,d). For $t \geq 0$, the external field is shifted toward well 1 in (a, c) and toward well 2 in (b, d). 
}
\label{fig:Cdipl}
\end{figure}

At \( t = \tau \), the coherent state generated in the subsystem of sites 1 and 2 is transferred to the subsystem of sites 1–3 or 2–3 (up to a global phase):  
\begin{eqnarray}
|\Psi_1^{\rm II}(\tau)\rangle &=& |\alpha, \beta\, e^{-i\phi}\rangle\rangle_{1,3} |0, n_4\rangle_{2,4},\nonumber\\
|\Psi_2^{\rm II}(\tau)\rangle &=& |\alpha, \beta\, e^{i\phi}\rangle\rangle_{2,3} |0, n_4\rangle_{1,4},\nonumber
\end{eqnarray}
where \( \phi = 3\nu\tau + \pi/2 \). The sequence of states generated at each step is schematically represented below: 

\[
\begin{array}{l l l}
\hline
\quad  t = -q\tau & \quad t = 0 &  \qquad t =\tau \\
\hline\\[-2mm]
& & \hspace{-.45cm}{\displaystyle |\alpha, \beta\, e^{-i\phi}\rangle\rangle_{1,3} |0, n_4\rangle_{2,4}} \\[1mm]
& & \hspace{-1.5cm}\tikz \draw[->] (0,.0) -- (0.5,.5) node[pos=0.6, left, xshift=-1mm, font=\scriptsize, text=black] {{\it k}=1};\\[-1mm]
|n_1,0,0,n_4\rangle \xrightarrow{k=3} & |\alpha,\beta\rangle\rangle_{1,2} |0,n_4\rangle_{3,4} & \\[1mm]
& & \hspace{-1.5cm}\tikz\draw[->] (0,0) -- (0.5,-0.5) node[pos=0.6, left, xshift=-1mm, ] {\scriptsize{\it k}=2}; \\[-1mm]
& & \hspace{-.45cm}{\displaystyle |\alpha, \beta\, e^{i\phi}\rangle\rangle_{2,3} |0, n_4\rangle_{1,4}}
\end{array}
\]

Figure~\ref{fig:Cdipl} shows a comparison between the numerical simulation and expression \eqref{ncdipl} for $q = 1/4$ (Figs. \ref{fig:Cdipl}(a) and \ref{fig:Cdipl}(b)) and $q = 2/3$ (Figs. \ref{fig:Cdipl}(c) and \ref{fig:Cdipl}(d)). 
These results demonstrate that the amplitude of the fractional population can be tuned throughout the time evolution 
under the action of $F_3$, during the first step of the protocol (shaded regions in Fig. \ref{fig:Cdipl}).
For $t\geq 0$, Figures \ref{fig:Cdipl}(a) and \ref{fig:Cdipl}(c) illustrate the field $F_1$ driving the atomic flow through the tunnelling channel $C_{2,3}$. Figures \ref{fig:Cdipl}(b) and \ref{fig:Cdipl}(d), on the other hand, show the flow through the $C_{1,3}$ channel resulting from the action of the field $F_2$. The oscillation period in Figs. \ref{fig:Cdipl}(b) and \ref{fig:Cdipl}(d) is longer than in Figs. \ref{fig:Cdipl}(a) and \ref{fig:Cdipl}(c) 
due to the fact that the values of $\sigma$ are closer to the critical values $\sigma_{\rm crit}$ in the latter cases. 

These results indicate that the protocol not only controls the direction and amplitude of the atomic flow, but also allows for the independent adjustment of the oscillation frequency via the intensity of the external field, effectively emulating 2:1 multiplexing and 1:2 demultiplexing operations. This integrated control mechanism over the dynamics of atomic populations substantially expands the operational range for possible atomtronic applications.


\section{Conclusions}
We have investigated ultracold dipolar atoms in a four-well star-shaped potential and identified the conditions for integrability, allowing an effective description of population dynamics in the resonant tunneling regime and providing an ideal scenario for exploring control mechanisms via an external field. We demonstrated that the oscillation frequency can be tuned by the field and classified atom routing operations using different initial Fock states. By introducing an additional step to adjust the oscillation amplitude and integrating these control mechanisms, we implemented atomic 2:1 multiplexing and 1:2 demultiplexing protocols. Our study identifies control mechanisms that could be useful for transporting and encoding information in a four-well system. This provides a fundamental building block based on ultracold dipolar atoms, that could contribute to advancements in quantum technologies, including atomtronics.

\begin{acknowledgments}
A.F. acknowledges support from CNPq - Edital Universal 406563/2021-7.  
K.W.W, and A.F. are grateful to the Brazilian agency CNPq (Conselho Nacional de Desenvolvimento Cient\'ifico e Tecnol\'ogico) for partial financial support. 
K.W.W. and A.F. were supported by the 
State of Rio Grande do Sul through FAPERGS - Edital FAPERGS/CNPq 07/2022 - Programa de Apoio à Fixação de Jovens Doutores no Brasil, contract 23/2551-0001836-5. We extend our deepest gratitude to Dr. Jon Links for his valuable and insightful contributions.
\end{acknowledgments}


\appendix


\section{Hamiltonian of the system}\label{app-Hamiltonian}
Here, we examine in detail the conditions to obtain the Hamiltonian \eqref{h} of the system.  

We consider $N$ dipolar bosons confined within a four-well potential arranged in a star configuration, represented by 
\begin{eqnarray}
V_{\rm opt} = V_{\rm trap} + V_{\rm ext}, \label{V_opt}
\end{eqnarray}
where $V_{\rm trap}$ represents the potential of a hexagonal lattice from which a plaquette containing four wells can be isolated using an external field associated with the potential $V_{\rm ext}$. In this configuration, the system is described by the extended Bose-Hubbard Hamiltonian, given by
\begin{align}
H =& \frac{U_0}{2}\sum_{j=1}^4N_i(N_i-1) +\frac{1}{2}\sum_{\substack{i,j=1\\i\neq j}}^4 U_{ij}N_iN_j + \sum_{i=1}^4 \sigma_i N_i\nonumber\\
   &-J [a_4^\dagger(a_1+a_2+a_3)+(a_1^\dagger+a_2^\dagger +a_3^\dagger)a_4],\nonumber
\end{align}   
where $a_i\,(a_i^\dagger)$ is the annihilation (creation) bosonic operator, and $N_i$ is the number operator of site $i$. $U_0$ and $U_{ij}$ are the on-site and inter-site interaction energies, respectively. The coupling $J$ is the hopping rate between nearest-neighbor sites, and $\sigma_i$ characterizes the strength of external field on site $i$. We assume that the energies involved do not excite the second Bloch band, so that the Wannier function in the lowest-band approximation can be represented by the Gaussian $\varphi_i({\bf r}) = \varphi({\bf r}-{\bf r}_i)$, where ${\bf r}_i$ is the center of site $i=1,2,3,4$. Under this assumption, the parameters are given by~\cite{Baranov2008}
\begin{eqnarray}
U_0&=& U_{\rm c}+U_{\rm dip},\nonumber\\
U_{\rm c} &=& g\int d{\bf r}\, |\varphi_1({\bf r})|^4,\nonumber\\
U_{\rm dip} &=& \int d{\bf r} \, d{\bf r}'\,|\varphi_i({\bf r})|^2V_{\text{DDI}}({\bf r}-{\bf r}')|\varphi_i({\bf r}')|^2,\nonumber\\
U_{ij} &=& \int d{\bf r} \, d{\bf r}'\,|\varphi_i({\bf r})|^2V_{\text{DDI}}({\bf r}-{\bf r}')|\varphi_j({\bf r}')|^2,\nonumber\\
J&=& -\int d{\bf r}\,\varphi_1({\bf r})\left[-\frac{\hbar^2}{2m}\nabla^2+V_{\text{trap}}({\bf r})\right]\varphi_4({\bf r}),\nonumber\\
\sigma_i&=& \int d{\bf r}\, |\varphi_i({\bf r})|^2V_{\text{ext}}({\bf r}),\nonumber
\end{eqnarray}
where the coupling $g= 4\pi \hbar^2 a /m$ characterizes the contact on-site interaction, $m$ is the mass of atom considered and the scattering length $a$ is controlled via Feshbach resonances. The inter-site interaction energy is determined by the dipole-dipole interaction (DDI) by means of the potential
\begin{eqnarray}
V_{\text{DDI}}({\bf r}) =\frac{\mu_0\mu_{\text{d}}^2}{4\pi}\frac{(1-3\cos^2\theta_P)}{|{\bf r}|^3},\nonumber 
\end{eqnarray}  
where $\mu_0$ is the vacuum magnetic permeability, $\mu_{\text{d}}$ is the permanent magnetic dipole moment of an atom and $\theta_P$ is the angle between the direction of polarization and the relative position of the particles. The on-site energy $U_{\rm dip}$ results from DDI between atoms within the same site. We are assuming that all atoms are polarized along the $z$-direction, orthogonally to the arrangement of the wells in the $xy$ plane.

Next, by using conservation of the total number of atoms $N=N_1+N_2+N_3+N_4$ and the identities
\begin{align}
\sum_{i=1}^4\frac{N_i(N_i-1)}{2} =& \frac{N(N-1)}{2} - N_4(N_1+N_2+N_3) \nonumber\\
                                &-(N_1N_2+N_1N_3+N_2N_3),\nonumber\\
N_4(N_1+N_2+N_3) = &\frac{1}{4}[N^2-(N_1+N_2+N_3-N_4)^2],\nonumber
\end{align}
we can reduce the Hamiltonian (up to a global constant $U_0N(N-1)/2 -UN^2+(\sigma+\sigma_4)N$) to
\begin{align}
H =& U(N_1+N_2+N_3-N_4)^2 +\sigma(N_1+N_2+N_3-N_4)   \nonumber\\
   & +(U_{12}-U_0)(N_1N_2+N_1N_3+N_2N_3)\nonumber\\
   & -J [a_4^\dagger(a_1+a_2+a_3)+(a_1^\dagger+a_2^\dagger +a_3^\dagger)a_4],\nonumber
\end{align}  
where we define the parameter
\begin{eqnarray}
U= \frac{U_0-U_{14}}{4},\nonumber
\end{eqnarray}
and we set the parameters $\sigma_1=\sigma_2=\sigma_3=2\sigma+\sigma_4$ in the case where the external field is aligned with the center of the system. The integrability is achieved when $U_{12}=U_0$, which leads to the Hamiltonian \eqref{h}.


\section{Experimental feasibility}\label{app-Experimental}
In this appendix, we discuss a possible experimental setup for implementing the model in the laboratory.

The first part of~\eqref{V_opt} represents the potential of the hexagonal optical lattice that can be obtained by interfering three co-planar counter-propagating standing waves with wavelength $\lambda$ on the $xy$ plane and an additional vertical counter-propagating stand wave in the $z$-direction used to control the aspect ratio of the potential trap. This potential is given by 
\begin{eqnarray}
V_{\text{trap}} = V_0\sum_{i=1}^3\cos^2(k\,{\bf r}\cdot {\bf u}_i+\Phi_i) + \frac{1}{2}m\omega_z^2 z^2, 
\end{eqnarray}
where ${\bf r} = (x,y)$, $k=\frac{2\pi}{\lambda}$ is the wave number and $V_0$ is the potential depth. 
The aspect ratio of the potential trap can be controlled by the potential depth $V_1$ of vertical counter-propagating stand-wave, which provides the trap frequency of $z$ direction 
\begin{eqnarray}
\omega_z=\sqrt{\frac{2V_1k^2}{m}}.\nonumber
\end{eqnarray}

The unit vectors are given by
\begin{eqnarray}
{\bf u}_1 = (\sqrt{3},\,1)/2,\; {\bf u}_2 = (-\sqrt{3},\,1)/2, \; {\bf u}_3={\bf u}_1+{\bf u}_2,\nonumber
\end{eqnarray}
and the phase 
\begin{eqnarray}
\Phi_i= (0,\,2\pi/3)\cdot {\bf u}_i,\nonumber
\end{eqnarray}
puts the center of the system at the origin such that the positions of sites are given by
\begin{eqnarray}
(x_1,y_1) &=& (0, l), \quad (x_2,y_2) =l(-\sqrt{3},\, -1)/2, \nonumber\\
(x_3,y_3) &=& l(\sqrt{3},\, -1)/2, \quad (x_4,y_4) = \left(0,0\right),\nonumber
\end{eqnarray} 
where $l = \lambda/3$ is the distance between neighboring sites. 

\begin{figure}
    \centering
 \includegraphics[width=1\linewidth]{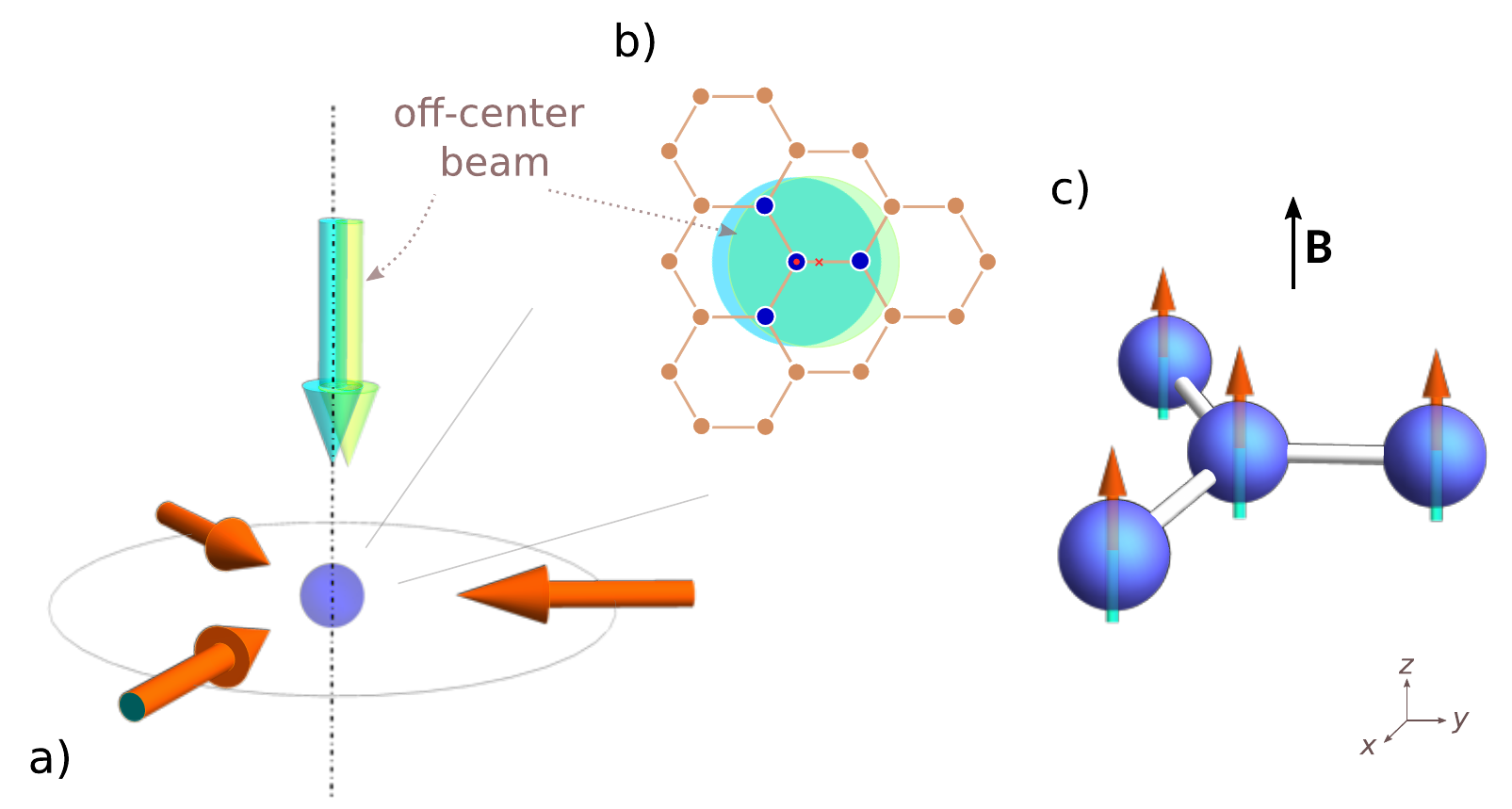}
   \caption{Schematic representation of the experimental setup. 
   (a) Three counter-propagating beams (orange arrows) generate a hexagonal lattice (blue sphere). 
   The cyan vertical arrow inserts a symmetry break when moved to one of the wells, as indicated by the yellow vertical arrow.
   (b) Two-dimensional schematic of the hexagonal lattice, highlighting four wells (dark blue dots) arranged in a star configuration, enlarged in 
   (c), where the arrows indicate the orientation dipoles induced by a magnetic field {\bf B}.}
    \label{fig:scheme}
\end{figure}

The second part of~\eqref{V_opt} corresponds to the potential of an external field, resulting from the superposition of a second weaker-intensity honeycomb lattice and a vertical Gaussian beam with a waist of $w$. The second lattice is arranged to form a potential barrier at the central well, creating a gradient of energy between well 4 and the edge wells. 
Alternatively, the Gaussian beam can be slightly offset from the center to generate an energy gradient in the outer wells as illustrated in Fig. \ref{fig:scheme}.
This setting allows the Hamiltonian parameters resulting from the external field to be manipulated independently over a wide range of values. The potential generated is given by 
\begin{eqnarray}
V_{\text{ext}} = V_2\sum_{i=1}^3\cos^2(k\,{\bf r}\cdot {\bf u}_i)+\frac{2V_3}{w^2}[(x-\Delta x)^2+(y-\Delta y)^2],\nonumber\\
\end{eqnarray}
where $V_2$ and $V_3$ are the potential depths, $\Delta x$ and $\Delta y$ represent the displacements along the $x$ and $y$ directions, respectively.  
The harmonic approximation of the potential trap at site $i$ is given by
\begin{eqnarray}
V_{\text{trap}}^{(i)} = \frac{1}{2}m\omega^2[(x-x_i)^2+(y-y_i)^2] + \frac{1}{2}m\omega_z^2 z^2,\nonumber
\end{eqnarray}  
where the trap frequency of $xy$ plane is given by
\begin{eqnarray}
\omega=\omega_x =\omega_y =\frac{\pi}{l}\sqrt{\frac{2V_0}{3m}},\nonumber
\end{eqnarray}
and the aspect ratio of potential trap is determined by the ratio $\kappa^2 =\omega_z/\omega$. This approximation determines the Wannier function of the lowest band Bloch of site $i$ represented by the Gaussian $\varphi_i({\bf r}) = \varphi({\bf r}-{\bf r}_i)$ where
\begin{equation}
\varphi({\bf r})=\prod_{s=x,y,z}(2\eta_s/\pi)^{1/4}e^{-\eta_s s^2}, \quad \eta_s =\frac{m\omega_s}{2\hbar}.\nonumber 
\end{equation}
The above wave function allows us to calculate the energy contribution due to the external field, which is given by:
\begin{eqnarray}
H_{\text{ext}} = \sum_{i=1}^4 \sigma_i N_i,\nonumber
\end{eqnarray}
where
\begin{eqnarray}
\sigma_i = \int d{\bf r}\, |\varphi_i({\bf r})|^2V_{\text{ext}}({\bf r}).\nonumber
\end{eqnarray}
When the Gaussian laser beam is displaced from the center toward site $k=1,2,3$ by $\Delta l$, the corresponding contribution to the Hamiltonian is given by 
\begin{eqnarray}
H_{\text{ext},k} &=& \sigma(N_1+N_2+N_3-N_4)\nonumber\\
&&+\nu(N_i+N_j-2N_k) +\Lambda(\Delta l)N, \quad (i\neq j\neq k). \nonumber 
\end{eqnarray}
where
\begin{eqnarray}
\sigma&=&-\frac{9}{8}e^{-\frac{2\pi^2}{9l^2\eta}}V_2+\frac{l^2}{w^2}V_3,\qquad \eta = \eta_x=\eta_y\nonumber\\
\nu&=&\frac{2l\,\Delta l}{w^2}V_3,\nonumber\\
\Lambda(\Delta l) &=& \frac{3}{8}\left(4+e^{-\frac{2\pi^2}{9l^2\eta}}\right)V_2+\frac{V_3}{w^2}\left[l^2+2(\Delta l)^2+\frac{1}{\eta}\right], \nonumber\\
\end{eqnarray}


Table~\ref{tab} provides the numerical values of the experimental parameters used throughout the text for the numerical simulations, to illustrate our results.

\begin{table}[h!]
\centering
\caption{Experimental values for $N=16$, $\Delta l = -l/2$, and $V_1/V_0 = 0.75$, $V_2/V_0=0.0004$, $V_3/V_0 =0.005$, where $h$ is the Planck constant, $a_0$ is the Bohr radius, $E_r=h^2 /(2m\lambda^2)$ is the recoil energy, and $m$ is the atomic mass.}
\vspace{0.3cm}
\begin{tabular}{lcc}
\hline
 Parameters&Symbols& Values \\
\hline
wavelength &$\lambda$ & 1064 nm\\
dipolar length & $a_{\rm dd}$ & 131.97 $a_0$\\
potential depth & $V_0$ & 70.96 $E_r$\\
interaction energy & $U/J$ & 0.51\\
aspect ratio & $\kappa $ & $1$ \\
trap frequency & $\omega/(2\pi)$ & $15.62$ kHz\\
waist of gaussian beam & $w$ & 2 $\mu$m\\
chemical potential & $\sigma/J$ & 1.56\\
energy offset & $\nu/J$ & 1.05 \\
\hline
\label{tab}
\end{tabular}\\
\end{table}

\section{Effective Hamiltonian}\label{app-Heff}
Here, we show that the effective Hamiltonian can be expressed in terms of conserved charges.
In the resonant tunneling regime, the interaction and external field terms, given by
\begin{equation}  
    H_0 = U(N_1 + N_2 + N_3 - N_4)^2 + \sigma(N_1 + N_2 + N_3 - N_4), \nonumber  
\end{equation}  
dominate over the tunneling part
\begin{equation}  
    V = -J \left[(a_1^{\dagger} + a_2^{\dagger} + a_3^{\dagger}) a_4 + a_4^{\dagger} (a_1 + a_2 + a_3) \right], \nonumber 
\end{equation} 
allowing $V$ to be treated as a perturbation in the Hamiltonian $H=H_0+V$, as given in \eqref{h}.
This perturbation allows us to calculate the transition rate $W = {2\pi} |T_{i\to f}|^2\delta(E_k-E_s)$, for a second-order transition in the form $|i\rangle \to |s\rangle \to |f\rangle$, where 
\begin{equation}  
   T_{i\to f} = \sum_{s } \frac{\langle f|V|s\rangle \langle s| V|i\rangle}{E_i - E_s},\nonumber  
\end{equation}  
and $E_{i(s)}$ is an eigenvalue of $H_0$. 
For instance, the direct transition \( |i\rangle = |n_1,n_2, n_3,n_4\rangle \to |f\rangle =|n_1-1,n_2+1, n_3,n_4\rangle \) is suppressed but occurs via virtual intermediate states \( |s\rangle \), such as
\begin{eqnarray}
&&|i\rangle \to |n_1-1,n_2, n_3,n_4+1\rangle \to |f\rangle,\nonumber\\
&&|i\rangle \to |n_1,n_2+1, n_3,n_4-1\rangle \to |f\rangle.\nonumber
\end{eqnarray}
Then, the identification of $T_{i\to f}$ via the element matrix of tunneling term $\langle f |\mathcal{J}_{\rm eff} \tau_{12} |i\rangle$  ($\tau_{ij}\equiv a_i^\dagger a_j+a_j^\dagger a_i$)  allows us to determine the effective hopping rate $\mathcal{J}_{\rm eff} = \mathcal{J}(0)$, where $\mathcal{J}(x)$ is defined in \eqref{Jx}. By symmetry arguments, other transitions follow by permuting the indices, leading to the effective Hamiltonian
\begin{eqnarray}
H^{\rm eff} &=&\mathcal{J}_{\rm eff}(\tau_{12}+\tau_{23}+\tau_{13})\nonumber\\
&=&  \mathcal{J}_{\rm eff}[2(N_1+N_2+N_3)-3(Q+\widetilde{Q})], \label{H_ij}  
\end{eqnarray}
where $Q=u^\dagger u$, $\widetilde{Q}=v^\dagger v$,  $u = (a_1 - a_2) / \sqrt{2}$, $v = (a_1 + a_2 - 2a_3) / \sqrt{6}$. Since $H^{\rm eff}/\mathcal{J}_{\rm eff}$ is a sum of particle-number operators, its eigenvalues are integers, indicating that the energy levels of the effective Hamiltonian are uniformly spaced.  
For the case of broken symmetry caused by the field $F_k$, the effective Hamiltonian is given by
\begin{eqnarray}
H_k^{\rm eff} &=& \mathcal{J}(\nu)[2(N_1+N_2+N_3)-3(Q+\widetilde{Q})]\nonumber\\
&&+\nu (N_i + N_j - 2N_k), \quad (i\neq j\neq k).\nonumber
\end{eqnarray}
For $\nu\gg \mathcal{J}(\nu)$, the external field, e.g., $F_3$, induces transitions of order $O[(\mathcal{J}(\nu))^2]$ of the form $|i\rangle \to |n_1-1,n_2,n_3+1,n_4\rangle \to |f\rangle$ and $|i\rangle \to |n_1-1,n_2+1,n_3-1,n_4\rangle \to |f\rangle$, so the part $ \mathcal{J}(\nu)(\tau_{13}+\tau_{23}) + \nu (N_1 + N_2 - 2N_3)$ of $H_3^{\rm eff}$ contributes effectively as $(\mathcal{J}(\nu))^2 \tau_{12}/(3\nu)$, and the dynamics remains confined to the subsystem of wells 1 and 2. Thus, in this regime, the effective Hamiltonian reduces, in general, to:
\begin{eqnarray}
H^{\text{eff}}_{k} &=&\zeta \tau_{12}+ \nu(N_i + N_j - 2N_k), \qquad (i \neq j \neq k)\nonumber\\
&=&\zeta (N_i + N_j - 2 Q_k)+ \nu(N_i + N_j - 2N_k), \label{heffi}
\end{eqnarray}
where $\zeta$ is given in \eqref{zeta}, and 
\begin{eqnarray}
Q_{k} = \frac{1}{2} (N_i + N_j - a_i^\dagger a_j - a_j^\dagger a_i), \qquad (i \neq j \neq k), \nonumber 
\end{eqnarray}  
is the conserved operator.


\section{Derivation of expectation values}\label{app-ExpValue}
In this section, we provide the main steps to derive the expectation value of populations from the effective Hamiltonians $\mathcal{H} = H^{\rm eff}$ and $\mathcal{H} = H_k^{\rm eff}$, given in \eqref{heff} and \eqref{heffi}, respectively. For simplicity, we consider here only $\mathcal{H} = H_3^{\rm eff}$; the other cases can be obtained analogously. The effective Hamiltonians can be written in a general form
\begin{eqnarray}
\mathcal{H} = \sum_{j,k=1}^3 c_{jk}a_j^\dagger a_k,\nonumber
\end{eqnarray}
where $c_{j,k}$ are the elements of the matrix
\begin{eqnarray}
c = \begin{cases}\left(\begin{array}{ccc}0&\mathcal{J}_{\rm eff}&\mathcal{J}_{\rm eff}\\
\mathcal{J}_{\rm eff}&0&\mathcal{J}_{\rm eff}\\
\mathcal{J}_{\rm eff}&\mathcal{J}_{\rm eff}&0
\end{array}\right) &,\text{ if } \mathcal{H} = H^{\rm eff},  
\\
 \left(\begin{array}{ccc}\nu&\zeta &0\\
\zeta &\nu&0\\
0&0&-2\nu
\end{array}\right)&,\text{ if }\mathcal{H} = H_3^{\rm eff}
\end{cases}.\nonumber
\end{eqnarray}
In each case, we consider the transformation of the basis operator in the form $b = M a$, where $a$ and $b$ represent the vectors 
\begin{eqnarray}
a = \left(\begin{array}{c}a_1\\a_2\\a_3\end{array}\right), \qquad b = \left(\begin{array}{c}b_1\\b_2\\b_3\end{array}\right), \nonumber  
\end{eqnarray}      
and 
\begin{eqnarray}
M = \begin{cases}\left(\begin{array}{ccc}\frac{1}{\sqrt{2}}&-\frac{1}{\sqrt{2}}&0\\
\frac{1}{\sqrt{3}}&\frac{1}{\sqrt{3}}&\frac{1}{\sqrt{3}}\\
\frac{1}{\sqrt{6}}&\frac{1}{\sqrt{6}}&-\frac{2}{\sqrt{6}}
\end{array}\right) &,\text{ if } \mathcal{H} = H^{\rm eff}  
\\
 \left(\begin{array}{ccc}\frac{1}{\sqrt{2}}&\frac{1}{\sqrt{2}}&0\\
\frac{1}{\sqrt{2}}&-\frac{1}{\sqrt{2}}&0\\
0&0&1
\end{array}\right)&,\text{ if }\mathcal{H} = H_3^{\rm eff}
\end{cases},\nonumber
\end{eqnarray}
whose inverse is given by $M^{-1}=M^T$. 
Using the transformation, the Hamiltonian can be reduced to diagonal form
\begin{eqnarray}
\mathcal{H} = \Omega_1 N_1^b+\Omega_2 N_2^b+\Omega_3 N_3^b,\nonumber
\end{eqnarray}  
where $N_k^b = b_k^\dagger b_k$, and
\begin{eqnarray}
(\Omega_1,\,\Omega_2,\, \Omega_3) =\begin{cases}
(-\mathcal{J}_{\rm eff},\,2\mathcal{J}_{\rm eff},\,-\mathcal{J}_{\rm eff})&,\text{ if } \mathcal{H} = H^{\rm eff}\\
(\nu+\zeta ,\,\nu-\zeta ,\,-2\nu)&,\text{ if }\mathcal{H} = H_3^{\rm eff} 
\end{cases}.\nonumber
\end{eqnarray}
In this form, we obtain
\begin{eqnarray}
e^{i t \mathcal{H}}a_je^{-i t \mathcal{H}} = \sum_{k=1}^3M_{kj}M_{kl}e^{i t \Omega_k}a_l.\nonumber
\end{eqnarray}
Then, using the above expression for the initial state $|\Psi_{\rm ini}\rangle = |n_1,n_2,n_3,n_4\rangle$, we obtain the expectation value
\begin{eqnarray}
\langle N_j\rangle &=& \langle \Psi_{\rm ini}| e^{i t \mathcal{H}}a_j^\dagger a_j e^{-i t \mathcal{H}}|\Psi_{\rm ini}\rangle\nonumber\\
&=& \sum_{k,l,q=1}^3 M_{kj}M_{kq} M_{lj}M_{lq}e^{i t (\Omega_l-\Omega_k)}n_q.\nonumber 
\end{eqnarray}
After some algebraic manipulations, the above expression leads to the formulas \eqref{n_k} and \eqref{njk2}.


\newpage

\bibliography{biblioCheck}

\end{document}